# How much could we cover a set by c.e sets? (1)

Farzad Didehvar

didehvar@aut.ac.ir

Mohsen Mansouri

m.mansouri@aut.ac.ir

Zahra Taheri

Zahra-taheri@aut.ac.ir

**Abstract**" How much c.e. sets could cover a given set?" in this paper we are going to answer this question. Also, in this approach some old concepts come into a new arrangement. The major goal of this article is to introduce an appropriate definition for this purpose.

**Introduction**

In Computability Theory (Recursion Theory) in the first step we wish to recognize the sets which could be enumerated by Turing machines (equivalently, algorithms) and in the next step we will compare these sets by some reasonable order (Like Turing degree). Also sometimes with some extra information (Oracles) a class of non c.e. sets show the same behavior as c.e. sets (Post hierarchy and related theorems). Here we try another approach: "Let A be an arbitrary set and we wish to recognize how much this set might be covered by a c.e. set?" Although in some sense this approach could be seen in some definitions of Recursion Theory, but at the best of our knowledge it didn't considered as an approach yet, even though it is able to shed a light on some subjects of Computability of sets. Defining this approach is not quite straightforward and there are some obstacles to define them. To overcome these difficulties we modify the definitions.

We have an alternative problem here when we consider recursive sets and not c.e. sets. In this case, the problem would be:

" Let A be an arbitrary set and we wish to know that how much this set might be covered by a recursive Set?" Here, we try the first definition and the first problem.

# 1. Preliminary Definitions and introducing a new concept of Maximality

Throughout this article, we consider the sets with finite difference as the same.

**Definition 1.1** let A be an arbitrary set. Inner maxima of A ($A^+$) is an c.e. subset of A such that for each c.e. set $A^+ \subset B \subset A$, $B \setminus A^+$ is finite, and outer maxima of A, ($A^-$) is an c.e. subset of complement of A which the difference set of $A^-$ with each c.e. subset contained it, is finite. In general, we define the maxima of A as ($A^+, A^-$).

Throughout this article we apply simply the concept of inner maxima.

**LEMMA 1.2:** Suppose an arbitrary set A which include simple set S. The same simple set is an inner maxima of A.

**Proof:** It is immediate.

**Definition 1.3** The set A is called inner maxima iff it has an inner maxima.

Figure 1:

## 2. The thin sets and some essential relations:

Here, our attempts are defining the first definitions in order to make a hierarchy between sets based on the proposed definition. In this section we will have more or less negative results to put forward our idea by using some natural definitions.

**Definition 2.1** A set which contains no infinite c.e. set is called a thin set.

**Lemma 2.2** The thin sets are closed under intersection and union.

**Lemma 2.3** Any c.e. set is inner maxima in the union of the set and a thin set.

**Lemma 2.4** Any inner maximal set in B is the difference set of B and a thin set.

**Theorem 2.5** At each Turing degree of computability we have a thin set.

**Proof** In each Turing degree of computability we have a simple set and the complement of a simple set is a thin set. □

**Definition 2.6** Consider a new relation on *A*, as follows:

$$A_j < A_i \leftrightarrow (A_j \subseteq A_i, (A_j - A_i) \text{ is not finite})$$

So we can consider the class A devised by<, (A, <) as a directed graph, i.e there is a directed edge from Aj to Ai if and only if Aj < Ai. This relation is transitive and anti-reflexive and does not have any cycle.

**Theorem 2.7** The graph of every no inner maxima set is unique, up to isomorphism.

**Proof.** By back and forth method … .

Consequently, there are exactly two cases about the set of "inner maxima of a set: if there is an inner maxima it is unique up to finite difference unless there are infinity many inner maxima. (The same for outer maxima)

In the second case, we call the set "embarrassed set".

We could generalize this concept to "embarrassed set respect to oracle X".

**Note:** The complement of halting set is an embarrassed set.

In following definition and theorem we conclude that the class $A := \{A_i : A_i \text{ is r.e}, A_i \subset A\}$) respect to the relation $\supset$ is somehow irregular, for embarrassed set A.

Here, we define an order between subsets of N:

**Definition 2.8** For any given subsets of N like A and B we have an order between the sets as following:

$A \gg B \leftrightarrow \exists T_1, T_2 \ (A-T_1) \supseteq (B - T_2)$

Here $T_1, T_2$ are Thin sets.

By this order, in a natural way we have the equivalence relation $\approx$ between subsets of natural numbers, (i.e. $A \approx B$ iff $A \gg B$ and $B \gg A$).

**Theorem 2.9** $A \approx B$ iff the difference set between A and B be a thin set.

**Theorem 2.10** Let A be a suset of B, and $C_{A,B} = \{X: A \subset X \subset B\}$, and define $c_{A,B}$ as $C_{A,B}/\approx$.

It is notable to know that, the class of all c.e sets under inclusion could be embedded in $c_{AB}$ iff ~(A ≈ B) (Equivalently, the difference set between A and B is not a c.e. set).

### 3. Roots of subsets of N:

Theorem 2.7 shows we are somehow failed to capture the ideas in the first of article, here we try to give another way.

**Definition 3.1.** Let we consider the set X as an oracle and $\{\varphi_i{}^X\}_{i \in N}$ as X-computable functions.

We define inner X-maxima sets in an analogous way we defined inner maxima set.

$$r(A) = \{X: A \text{ is an inner } X - maxima \text{ set}\}$$

**Lemma:** For any set B subset of N

$$B >_T A \rightarrow B \in r(A)$$

**Lemma 3.2.** There is an c.e. set A such that $|r(A)| = 2$.

We generalize the above lemma as following:

**Theorem 3.3** For any natural number n there is an c.e. set $A$ such that $|r(A)| = n$.

**Theorem 3.4** There is an c.e. set A such that $|r(A)| = \infty$.

**Definition 3.5** We define the relation R on subsets of N as following:

$$R(A,B) \leftrightarrow A \in r(B)$$

**Theorem 3.6:** R is not symmetric.

**Theorem 3.7:** R is not anti-symmetric.

**Definition 3.7:** We define the relation $R_1$ on subsets of N as following:

$$R_1(A, B) \leftrightarrow (R(A, B) \& R(B, A))$$

**Remark:** $R_1$ defines an equivalence relation on subsets of N.

**Definition 3.7:** We define the relation $R_2$ on subsets of N as following:

$$R_2(A, B) \leftrightarrow \exists X \, R_1(A, X) \& R(A, X)$$

**Theorem 3.8:** $R_2$ is a quasi order relation.

**Remark & Conclusion:** So we are able to define $R_2$-classes, and the ordering induces by $R_2$ on that.

**Acknowledgement**

The authors would like to thank Amirkabir University , also it is partially supported by IPM, 2009-2010. The ordering of names are in arrangement of new regulation in Amir -Kabir University.